\documentclass[sigconf]{acmart}

\AtBeginDocument{%
  \providecommand\BibTeX{{%
    \normalfont B\kern-0.5em{\scshape i\kern-0.25em b}\kern-0.8em\TeX}}}

\setcopyright{acmcopyright}
\copyrightyear{2018}
\acmYear{2018}
\acmDOI{XXXXXXX.XXXXXXX}

\acmConference[Conference acronym 'XX]{Make sure to enter the correct
  conference title from your rights confirmation emai}{June 03--05,
  2018}{Woodstock, NY}
\acmPrice{15.00}
\acmISBN{978-1-4503-XXXX-X/18/06}

\usepackage[inline]{enumitem}
\usepackage{layouts}

\begin{document}

\title[Nobody Wants to Work Anymore]{Nobody Wants to Work Anymore:\\An Analysis of r/antiwork and the Interplay between Social and Mainstream Media during the Great Resignation}

\author{Alan Medlar}
\affiliation{%
  \institution{University of Helsinki}
  \streetaddress{Helsinki}
  \country{Finland}
}
\email{alan.j.medlar@helsinki.fi}

\author{Yang Liu}
\affiliation{%
  \institution{University of Helsinki}
  \streetaddress{Helsinki}
  \country{Finland}
}
\email{yang.liu@helsinki.fi}

\author{Dorota G{\l}owacka}
\affiliation{%
  \institution{University of Helsinki}
  \streetaddress{Helsinki}
  \country{Finland}
}
\email{dorota.glowacka@helsinki.fi}


\begin{abstract}

r/antiwork is a Reddit community that focuses on the discussion of worker exploitation, labour rights and related left-wing political ideas (e.g.~universal basic income). In late 2021, r/antiwork became the fastest growing community on Reddit, coinciding with what the mainstream media began referring to as the Great Resignation. This same media coverage was attributed with popularising the subreddit and, therefore, accelerating its growth. In this article, we explore how the r/antiwork community was affected by the exponential increase in subscribers and the media coverage that chronicled its rise. We investigate how subreddit activity changed over time, the behaviour of heavy and light users, and how the topical nature of the discourse evolved with the influx of new subscribers. We report that, despite the continuing rise of subscribers well into 2022, activity on the subreddit collapsed after January 25th 2022, when a moderator's Fox news interview was widely criticised. While many users never commented again, longer running trends of users' posting and commenting behaviour did not change. Finally, while many users expressed their discontent at the changing nature of the subreddit as it became more popular, we found no evidence of major shifts in the topical content of discussion over the period studied, with the exception of the introduction of topics related to seasonal events (e.g.~holidays, such as Thanksgiving) and ongoing developments in the news (e.g.~working from home and the curtailing of reproductive rights in the United States).

\end{abstract}



\maketitle

\section{Introduction}

\begin{figure*}
    \centering
    \includegraphics{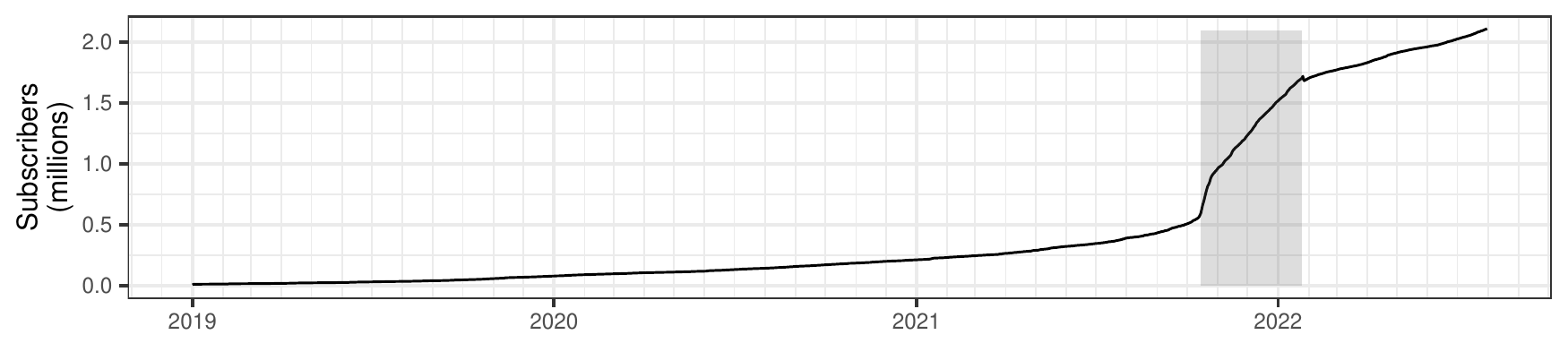}
    \caption{The number of subscribers to the r/antiwork subreddit from 2019 onwards. The grey box highlights the period from October 15 2021-January 25 2022.}
    \label{fig:subs}
\end{figure*}

In 2021, the attrition rate of employees in the global workforce 
reached record highs in an economic trend that became known as the Great Resignation\footnote{\url{https://www.bloomberg.com/news/articles/2021-05-10/quit-your-job-how-to-resign-after-covid-pandemic}}. The COVID-19 pandemic had caused many workers to leave the labour force because of problems related to child and social care arrangements, early retirement and even death \cite{fry2022resigned}. 
The resulting labour shortages led to wage growth, encouraging workers to quit their jobs and seek opportunities elsewhere \cite{parker2022majority}. 
More broadly, the trauma inflicted by the pandemic led many to question their relationship with work and to demand better working conditions \cite{sheather2021great, sull2022toxic}. The Great Resignation was widely reported on in the mainstream media, with coverage often linking to social media, 
e.g.~``{\em Man Quits His Job With Epic 'Have a Good Life' Text and People Are Impressed}''\footnote{\url{https://www.newsweek.com/1639419}}, 
``{\em Quitting Your Job Never Looked So Fun}''\footnote{\url{https://www.nytimes.com/2021/10/29/style/quit-your-job.html}} and 
``{\em Scroll through TikTok to see the real stars of the workplace}''\footnote{\url{https://www.ft.com/content/c7f8fb0e-8f1a-4829-b818-cb9fe90352fa}}.
Indeed, media articles often presented the growing popularity of r/antiwork\footnote{\url{https://www.reddit.com/r/antiwork/}}, a Reddit community, as emblematic of the significance of the Great Resignation\footnote{\url{https://www.ft.com/content/1270ee18-3ee0-4939-98a8-c4f40940e644}} (see Figure~\ref{fig:subs}). 

r/antiwork is a subreddit created to discuss worker exploitation, labour rights and the antiwork movement, irreverently encapsulated by the subreddit's slogan of ``{\em Unemployment for all, not just the rich!}''. 
Throughout the pandemic, r/antiwork enjoyed continuous subscriber growth, increasing from ~80,000 members at the start of 2020 to over 200,000 in less than a year. 
However, after becoming the subject of mainstream media coverage in mid-October 2021, the number of subscribers increased by over 330,000 within a two week period -- an increase of 57\% -- making it the fastest growing subreddit at the time (see grey region from Figure~\ref{fig:subs}). 
Interactions with the media continued to shape r/antiwork: Doreen Ford, a longtime moderator of the subreddit, was interviewed by Fox News on January 25 2022. The interview was controversial, resulting in the subreddit briefly going private, many members unsubscribing and a reduction in the rate of subscriber growth throughout 2022. 
Numerous redditors observed that there exists a tension between the moderators, who tend to hold more radical political views, and newer members of the subreddit who are more concerned with organised labour and reforming the current economic system\footnote{\url{https://www.reddit.com/r/SubredditDrama/comments/sdesxw/comment/huc9wf9/}}. Indeed, there are numerous posts from long-term members lamenting how the subreddit has changed over time, from discussing how ``{\em society would/could function without unnecessary labor}'' to users ``{\em posting real and fake text messages of quitting their job}''\footnote{\url{https://www.reddit.com/r/antiwork/comments/qfi56h/}}.

Reddit has been the subject of numerous studies on social media behaviour. 
These studies have shown that large numbers of new users can be disruptive to an online community \cite{kiene2016surviving}.
They can impact communication norms \cite{Haq2022short} and behave in ways that are harmful to the community \cite{kraut2012building}. 
However, even in extreme cases, like when a subreddit gets defaulted (made a default subreddit for newly registered Reddit accounts), 
the community can still remain high-quality and retain its core character \cite{lin2017better}.
Other studies have highlighted how the mainstream media can influence social media and the general public. 
For example, public attention of COVID-19 on Reddit was mainly driven by media coverage \cite{gozzi2020collective} and 
negative media articles led to numerous hateful subreddits being banned by Reddit, including r/TheFappening, r/CoonTown and r/jailbait. 
Media coverage has also been shown to have negative consequences on social media: it can increase problematic online behaviour \cite{Habib_Nithyanand_2022} and 
banning subreddits has increased hate speech elsewhere on Reddit \cite{horta2021platform}.
To our knowledge, however, there are no studies where a subreddit's rapid rise was so intertwined with media coverage and, moreover, 
where a media event was the catalyst in its decline. Furthermore, we are unaware of any other studies specifically related to r/antiwork.

To understand how r/antiwork was impacted by media events, we performed a quantitative analysis of over 300,000 posts and 12 million comments from January 2019 to July 2022. We performed a time series analysis of users posting and commenting behaviour, and investigated how user activity on r/antiwork was affected by the initial media articles in October 2021 and the Fox News interview in January 2022. Next, we categorised users as light and heavy users to understand how different types of user contribute to the subreddit. Lastly, we used topic modelling to understand whether the influx of new users had changed the discourse on r/antiwork, e.g.~focusing more on the topic of quitting their jobs rather than more serious topics related to the antiwork movement.
In summary, we ask the following research questions:

\begin{itemize}[leftmargin=4ex]
    \item {\bf RQ1 Subreddit Activity:} How did subreddit activity change after the increase in subscribers that coincided with coverage in the mainstream media?
    \item {\bf RQ2 User Types:} How was the posting and commenting behaviour of heavy and light users impacted by the growth in subscribers?
    \item {\bf RQ3 Content Analysis:} Did the influx of new users change the discourse in terms of the distribution of topics discussed?
\end{itemize}

In the remainder of this paper, we will answer these research questions and
discuss how our results relate to existing work on social media analysis.


\section{Related Work}

In this section, we briefly review the work in three research areas related to this article:
the impact of mainstream media on social media activity, massive growth in social media users, and previous analyses of Reddit data. 

\subsection{Impact of Mainstream Media on Social Media Activity}
Mainstream media coverage, such as newspaper articles, television programmes and radio broadcasts, on political or social topics can lead to increased awareness of that topic on social media, such as Reddit. Moreover, elevated interest in an event by mainstream media can impact the number of users as well as their activity on social media platforms. 
For example, Chew et al.~\cite{chew2010pandemics} and Tausczik et al.~\cite{tausczik2012public} examined the trajectories of activities on social media (Twitter and web blogs) during the H1N1 pandemic and noticed that peaks in user activity coincided with major news stories. Similarly, Gozzi et al.~\cite{gozzi2020collective} showed that during the COVID-19 pandemic, user activity on Reddit and searches on Wikipedia were mainly driven by mainstream media coverage.


The popularity of a topic in mainstream media can also lead to an increase in moderation activities on social media platforms, particularly on Reddit. For example, Reddit’s administrative interventions caused by violations of their content policy for toxic content occurred more frequently as a result of media pressure \cite{Habib_Nithyanand_2022}. Moreover, mainstream media attention on subreddits with toxic content further exacerbated the toxicity of their content \cite{Habib_Nithyanand_2022}. Horta Ribeiro et al.~\cite{horta2021platform} further studied user activity and content toxicity after r/The\_Donald and r/incels were banned due to media-driven moderation. They found a significant decrease in users' posting activity, but an increase in activities associated with toxicity and radicalization.

Our work is unique in that through the analysis of r/antiwork we study two simultaneous mainstream media-driven impacts on social media: 1) the massive growth in subscribers to r/antiwork coinciding with increased coverage of the Great Resignation by mainstream media, and 2) a spontaneous decrease in user activity triggered by a heavily criticised interview of an r/antiwork moderator on Fox News. To the best of our knowledge, this is the first quantitative study of the impact of spontaneous decreases on Reddit. Although previous studies examined the decreases in user activity after moderation \cite{horne2017impact, Habib_Nithyanand_2022}, these decreases were due to platform bans rather than spontaneous user behaviour.     

\subsection{Massive Growth in Social Media Users}
A topic often studied in social media is the growth in new users \cite{kraut2012building}.
Previous studies suggest that an influx of newcomers can cause online community disruption due to new users failing to adhere to community norms \cite{kraut2012building} or cause an information overload in a given online community \cite{jones2004information}. 
In recent years, several studies analyzed the impact of a massive growth of users on social media.
Kiene et al. \cite{kiene2016surviving} present a qualitative study of the massive growth of the subreddit r/NoSleep, demonstrating that the massive growth of the subreddit did not cause any major disruptions.
Lin et al.~ further showed that communities can remain high-quality and similar to their previous selves after the influx of new members \cite{lin2017better}. 
The work of Chan et al.~illustrates that a sudden spike in the number of users is a source of potential disruptions for an online community, however large communities are less impacted than smaller ones \cite{chan2022community}.
Additionally, Haq et al. \cite{Haq2022short} examine linguistic patterns on r/WallStreetBets, suggesting that writing style differs significantly between long-term users and new users resulting from a period of sudden growth. 
Our work studies the impact of the massive growth in the number of users on r/antiwork. Our analysis provides a new perspective on how the behaviour of different types of users (i.e.~heavy and light posters and commenters) are affected by subscriber growth.

\subsection{Social Meda Analysis of Reddit}
Reddit, as one of the most popular social media platforms, is widely used to study online communities and social phenomena. Many studies focus on the analysis of specific subreddits. 
Ammari et al.~\cite{ammari2018pseudonymous} analysed gender stereotypes on r/Daddit and r/Mommit, and Sepahpour et al.~\cite{sepahpour2022mothers} compared audience effects of r/Daddit and r/Mommit with r/Parenting.
Leavitta et al. \cite{leavitt2014upvoting} studied how the content of different topics on r/sandy, a subreddit dedicated to hurricane Sandy, changed over time.  
Horta Ribeiro et al.~\cite{horta2021platform} explored the impact of the ban of the subreddit r/The\_Donald on  user activity and content toxicity.
Haq et al.~ \cite{Haq2022short} focused on the impact of sudden community growth in r/WallStreetBets during the GameStop short squeeze in January 2021. 
Our work is somewhat similar to \cite{Haq2022short} in the sense that both works study the influence of massive growth in users caused by specific external events. However, we analyse changes in user behaviour and discussion topics, whereas Haq et al.~focus on the writing style of long-term and new users. 
Other studies of Reddit communities investigated community loyalty and successes \cite{grayson2018temporal, hamilton2017loyalty, cunha2019all}, topic popularity prediction \cite{adelani2020estimating}, and multi-community engagement \cite{tan2015all, hessel2016science}.

\section{Methodology}

\subsection{Data}

We downloaded all posts and comments on the r/antiwork subreddit from January 1 2019 to July 31 2022 using the PushShift API\footnote{\url{https://pushshift.io/}} \cite{baumgartner2020pushshift}. 
%
%
We only considered posts with at least one associated comment as a proxy for 
duplicate posts referencing the same event, 
off-topic and spam posts, 
as well as posts that received no user engagement for other reasons.
The data set contained 304,096 posts and 12,141,548 comments.
These posts were made by 119,746 users (posters) and the comments were made by 1,298,451 users (commenters).



We preprocessed the data set to remove comments that could potentially bias our analysis.
We filtered out comments that:
\begin{enumerate*}[label=(\roman*)]
    \item were removed by users or moderators, but remain in the data set as placeholders (comments are typically removed for violating community guidelines), or 
    \item were comments from bots (e.g.~the AutoModerator bot, or where the body of the comment began {\em ``I am a bot\dots''}, as many do by convention).
\end{enumerate*}
After filtering, 11,665,342 comments remained in the data set (96.1\%). 
We removed posts that had zero comments after filtering, leaving 284,449 posts (93.5\%)


%
%
%

\subsection{Definitions}

\subsubsection{User Types}
\label{sec:methods:users}
In our analysis, we compare the behaviour of two groups of users that we refer to as ``light'' and ``heavy'' users of r/antiwork. 
We define {\bf light posters or commenters} as those with only a single post or comment in the data set, respectively.
A majority of posters are light posters (75.1\%) and a high percentage of commenters are light commenters (42.5\%). 
We define {\bf heavy posters or commenters} as the top 1\% of users ranked in descending order by number of posts or comments, respectively. 
Overall, heavy posters made 10.1\% of posts and heavy commenters were responsible for 29.8\% of comments.



\begin{figure*}
    \centering
    \includegraphics{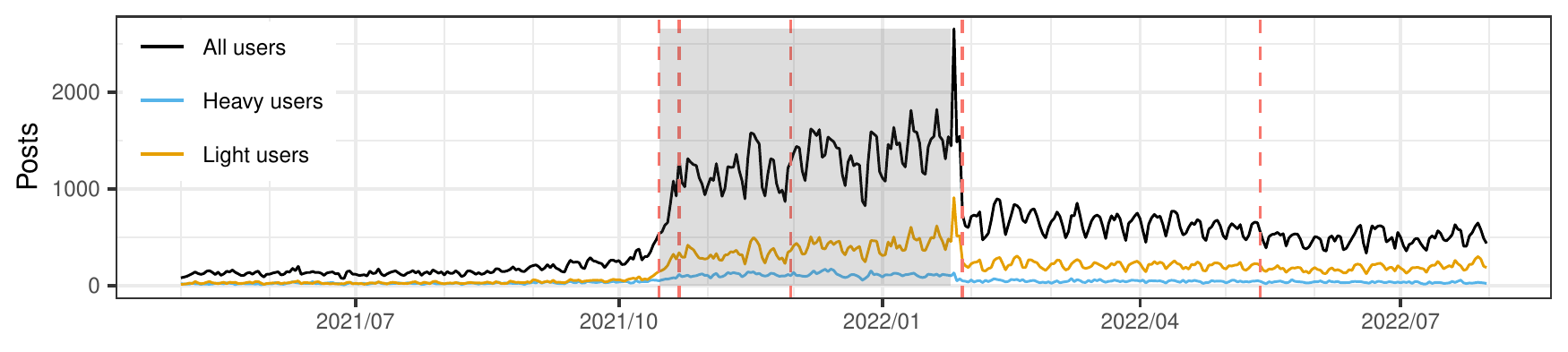}
    \caption{Total number of daily posts submitted to r/antiwork that received at least one comment. 
    A large proportion of posts (29.6\%) were made by light posters. Red dashed lines are results from change point detection.}
    \label{fig:posts}
\end{figure*}

\begin{figure*}
    \centering
    \includegraphics{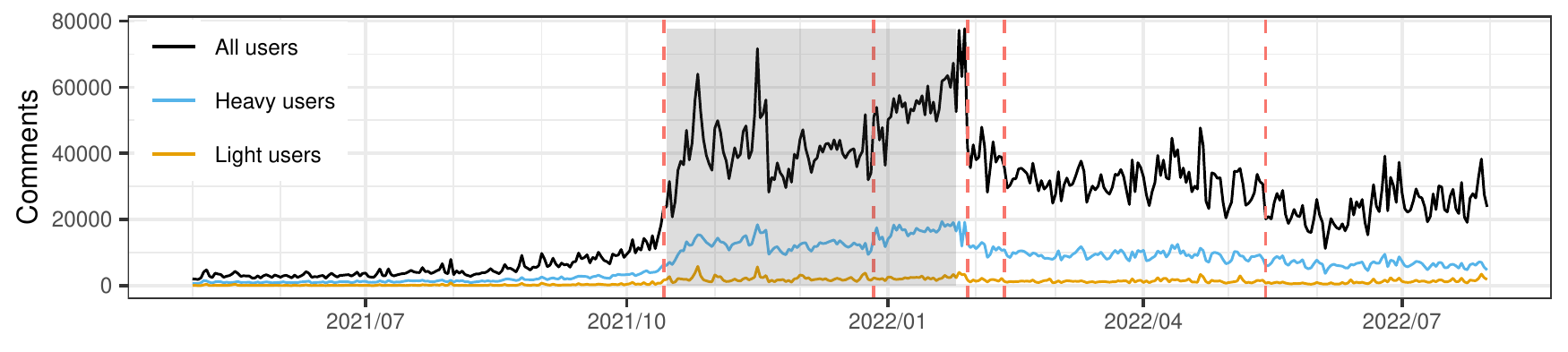}
    \caption{Total number of daily comments on r/antiwork. 
    A large proportion of comments (29.8\%) were made by heavy commenters. Red dashed lines are results from change point detection.}
    \label{fig:comments}
\end{figure*}

\subsubsection{Time Periods}
\label{sec:methods:timeperiod}
For our topic modeling analysis, 
we divided the data set into three time periods: 
\begin{itemize}
    \item {\bf Period 1:} January 1 2019--October 14 2021
    \item {\bf Period 2:} October 15 2021--January 24 2022
    \item {\bf Period 3:} January 25 2022--July 31 2022
\end{itemize}
These periods are delineated by two events in the mainstream media: 
the publication of a Newsweek article\footnote{\url{https://www.newsweek.com/1639419}}, 
which was the first example of a mainstream media article linking to a viral post\footnote{\url{https://www.reddit.com/r/antiwork/comments/q82vqk/}} on r/antiwork (October 15 2021) and 
the Fox News interview with Doreen Ford (January 25 2022). 
Period 2 is highlighted as a grey box in all figures where the $x$-axis represents time.

\subsection{Change Point Detection}

We use Classification And Regression Trees (CART) for change point detection \cite{breiman2017classification}. CART is a non-parametric method that uses a decision tree to recursively segment the predictor space into purer, more homogeneous intervals (often called ``splitting''). This segmentation process is terminated by a complexity parameter that regularises the cost of growing the tree by adding a penalty for adding additional partitions (``pruning''). In our case, we fit a regression tree with the dependent variable as the number of posts or comments, and the predictor space as each day from January 1 2019--July 31 2022. We used the rpart R package to create regression models \cite{therneau1997introduction}, the Gini index for splitting and a complexity parameter of 0.01 for pruning.

\subsection{Topic modelling} 
\label{sec:methods:topicmodel}

We use Latent Dirichlet Allocation (LDA) for topic modelling \cite{blei2003latent}. LDA is a generative model that defines a set of latent topics by estimating the document-topic and topic-word distributions within documents for a predefined number of topics. In our case, we consider each post to be a document and the contents of that document as the concatenation of all comments for that post. We do not include the post text as part of the document because a large proportion of post bodies are composed of images. 
We preprocessed comments for topic modelling by 
removing URLs and stop words, 
replacing accented characters with their ASCII equivalents, 
replacing contractions with their constituent words, and
lemmatizing all words. 
Finally, we filtered out posts with fewer than 50 comments 
leaving 11,368,863 comments (97.5\%) across 181,913 posts (64.0\%) for topic modelling.



LDA was applied to each of the three time periods separately (see Section~\ref{sec:methods:timeperiod}). 
Periods 1, 2 and 3 contained 40,794; 71,470 and 69,649 posts, respectively.
We evaluate the quality of topic models using the $C_{uci}$ coherence score \cite{newman2010automatic} to select the optimal number of topics. Each topic was labelled by a human annotator with knowledge of r/antiwork and topics were aligned between models using those labels and the Jensen-Shannon distance between topic-word distributions. Topic modelling was performed using the Gensim Python library \cite{rehurek_lrec}.

\section{Results}

In the following section, 
we characterise how posting and commenting activity changed during the period of increased media coverage (RQ1), 
then we investigate trends in the behaviour of heavy and light users (RQ2),
and, lastly, we see how the distribution of topics changed between the three time periods (RQ3).
Unless stated otherwise, all analyses refer to the time period between January 1 2019-July 31 2022. 
Figures are limited to the period between May 1 2021-July 31 2022 for the sake of clarity.

\subsection{RQ1: Subreddit Activity}

The mainstream media usually points to the number of r/antiwork subscribers to illustrate its growth and popularity (see Figure~\ref{fig:subs}). However, in addition to subscribing, users can interact with a subreddit by posting, 
commenting and voting.
As Reddit no longer provides the number of up- and down-votes to third parties, we focused on users' posting and commenting behaviour. 



\begin{figure*}
    \centering
    \includegraphics{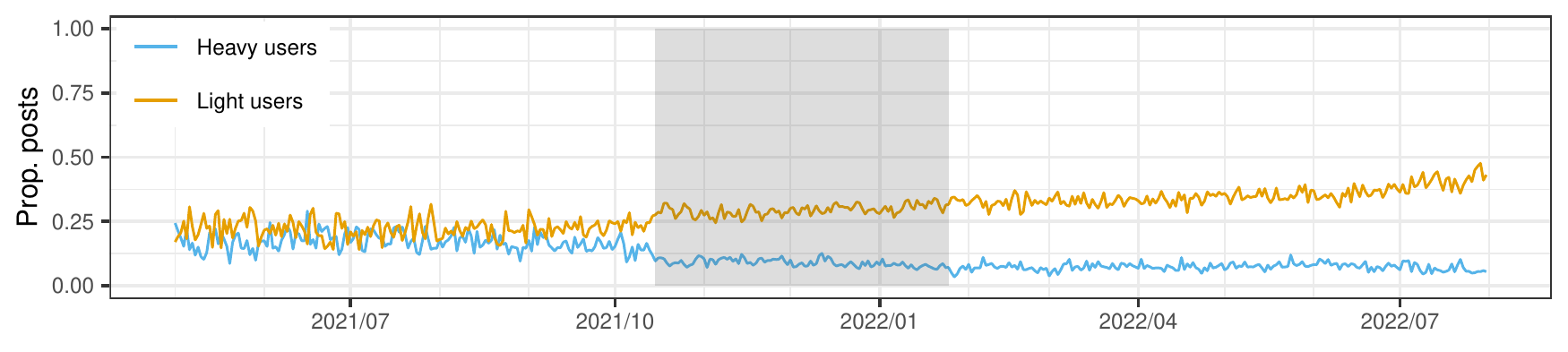}
    \caption{Proportion of posts from r/antiwork that received at least one comment 
    made by heavy and light posters. }
    \label{fig:postsprop}
\end{figure*}

\begin{figure*}
    \centering
    \includegraphics{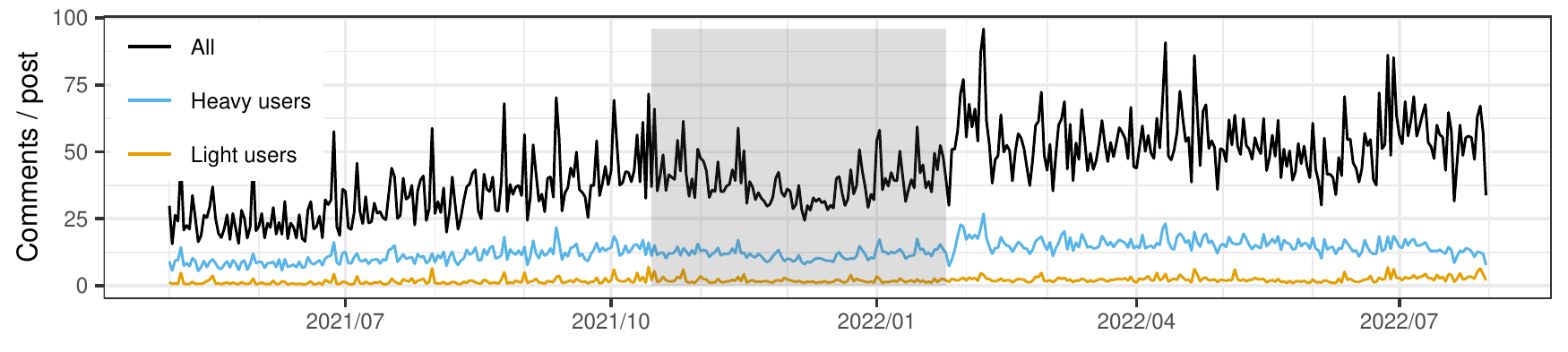}
    \caption{Average comments received by each post on r/antiwork. Average comments per post from heavy and light users remained relatively constant over time.}
    \label{fig:commentsperpost}
\end{figure*}

Figure~\ref{fig:posts} shows the daily number of posts submitted to r/antiwork that received at least one comment. 
Up until mid-2021, the average number of posts per day grew steadily, 
for example, increasing from 46.4 in January 2020, just prior to the start of the Coronavirus pandemic, to 76.8 in April 2021.
From May 2021, the rate of posting started to accelerate, consistently breaching 200 posts per day by September, before growing exponentially from October 9 to the weekend of October 23-24. 
From late October 2021, posting behaviour settled into a pattern of heightened activity during weekdays that dips during the weekends. 
At its peak, 2,658 posts were made on January 26, the day after Doreen Ford's Fox News interview, before collapsing to less than half the posting volume of the preceding month.
On January 27 2022, r/antiwork lost 38,228 subscribers (2.2\%) (see the right hand edge of the grey region in Figure~\ref{fig:subs}). For comparison, the second biggest dip in subscribers was on February 24 2019 when the number of subscribers decreased by 7.




Figure~\ref{fig:comments} shows similar trends in commenting behaviour: 
an exponential increase in mid-October 2021 followed by a sudden collapse in late January 2022. 
Unlike posting, however, there is no obvious differences between commenting volume on weekdays versus weekends. 
As with the posts to r/antiwork, the number of comments peaked during January 26-28, before falling 46.2\% on January 29 2022.



The dashed lines on Figures~\ref{fig:posts} and \ref{fig:comments} show the results from change point detection.
In Figure~\ref{fig:comments}, the first change on October 14 follows a viral post by u/hestolemysmile (the single most commented on post on r/antiwork\footnote{\url{https://www.reddit.com/r/antiwork/comments/q82vqk/}}).
In Figure~\ref{fig:posts}, the first two changes on October 15 and 22 coincide with the publication of widely-circulated articles by Newsweek and the New York Times, respectively. 
In both Figures~\ref{fig:posts} and \ref{fig:comments}, January 29 was identified as the number of posts and comments fell after the Fox News interview.
The remaining events appear to be around seasonal holidays: 
posts appear to increase following Thanksgiving (November 30), 
while comments increase on the first working day after Christmas (December 27).
The last events related to posting (May 13 2022) and commenting (February 11 and May 14 2022) do not appear to be related to specific events, but is the model acknowledging more gradual downward shifts in activity. 

\begin{figure*}
    \centering
    \includegraphics{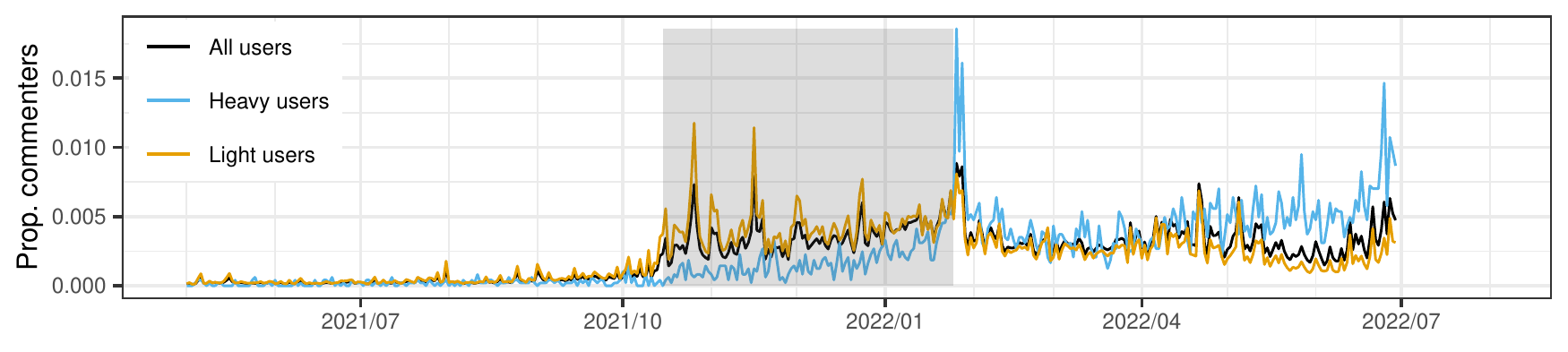}
    \caption{Proportion of users whose last comment to r/antiwork fell on each day. 
    Data from the last month in the data set was excluded as many of these users will continue commenting.}
    \label{fig:lastseen}
\end{figure*}

\begin{figure}
    \centering
    \includegraphics[width=\columnwidth]{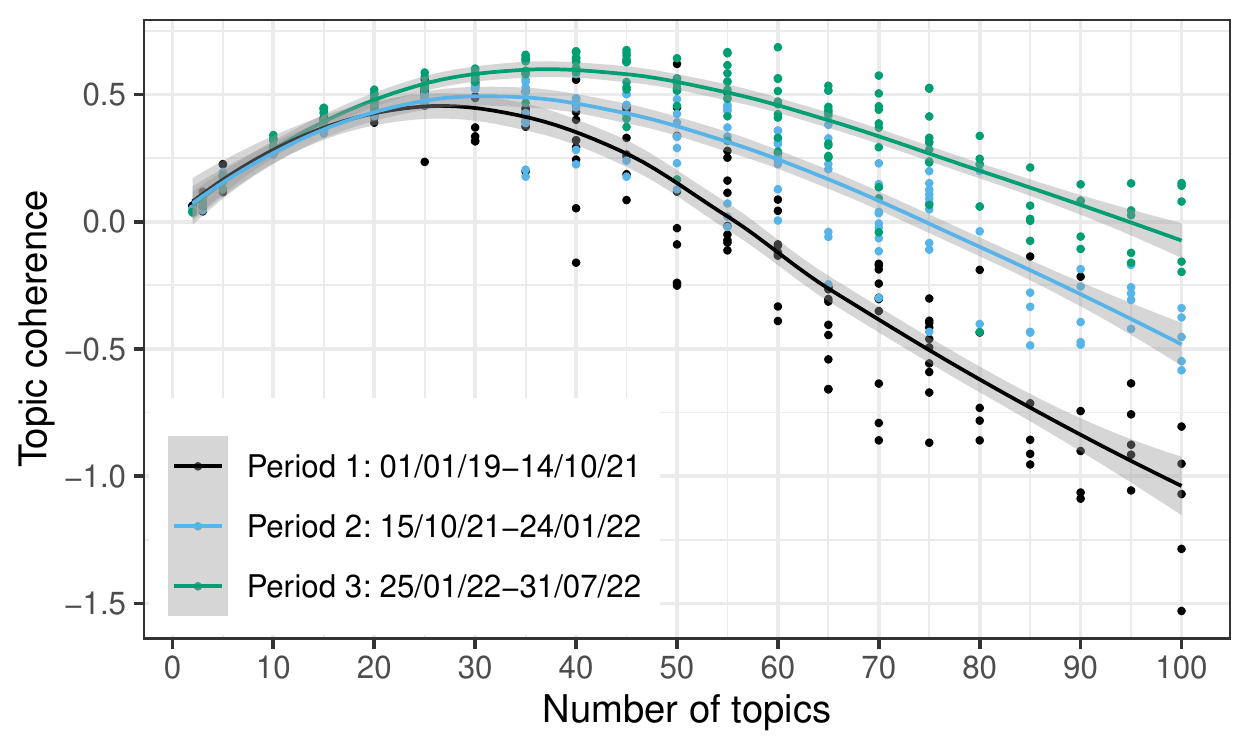}
    \caption{Topic coherence ($C_{uci}$) for different numbers of topics for three partitions of the data set.}
    \label{fig:coherence}
\end{figure}

\subsection{RQ2: Behaviour of Heavy and Light Users}

The results from RQ1 showed that consistently growing subscriber counts do not necessarily lead to ever-increasing numbers of posts and comments, but are contingent on external events. Here, we investigate the behaviour of heavy and light users (defined in Section~\ref{sec:methods:users}) to understand who is driving the changes in the volume of posts and comments. We also look at when users made their last comment to the subreddit to assess whether users stopped engaging with r/antiwork or simply comment less frequently after the interview on Fox News.


Figure~\ref{fig:posts} shows that posting behaviour is mostly driven by light posters, who were responsible for 29.6\% of posts, compared to 10.1\% for heavy posters. Figure~\ref{fig:postsprop} shows that the proportion of posts made by light and heavy posters were approximately equal prior to October 2021, but then start to diverge with almost half of posts coming from light posters by the end of July 2022.
%
%
Conversely, Figure~\ref{fig:comments} shows that heavy commenters make more comments in aggregate than light commenters (29.8\% vs. 4.7\%). Unlike users' posting behaviour, however, the average number of comments per post remained relatively constant over time for both types of commenters, a trend that appears to be unaffected by the surge in subscribers (see Figure~\ref{fig:commentsperpost}). 



Lastly, in Figure~\ref{fig:lastseen} we investigated when users made their last comment to r/antiwork (we omitted the last month's data for clarity as many of these users will continue commenting in the future). 
Between October 2021 and January 2022, a majority of users commenting for the last time were light commenters, i.e.~their last comment is their first and only comment. The proportion of heavy commenters making their last comment remained low until January 26-28 2022 when 4.4\% of heavy commenters made their final comment. After January 2022, it was equally likely that heavy and light commenters stopped commenting until May 2022 when it became more likely for heavy commenters to stop commenting than light commenters of r/antiwork.

\subsection{RQ3: Content Analysis}

In RQ1, we showed that the volume of posts and comments increased dramatically in October 2021 before collapsing in January 2022. 
In RQ2, however, we saw that an increasing proportion of posts came from light users, i.e.~users who only post once. 
We want to understand how these two phenomena affected what was discussed on r/antiwork using topic modelling.
We investigate the optimal number of topics and contrast the topic distributions for the three time periods defined in Section~\ref{sec:methods:timeperiod}.


We used topic coherence to identify the optimal number of topics. 
Figure~\ref{fig:coherence} shows the coherence scores for topic models with 5-100 topics in increments of 5. 
We performed either 5 or 10 replicates for each number of topics for each time period (more replicates were run for 15-75 topics where the coherence score was maximised). The optimal number of topics was 25, 30 and 40 for periods 1, 2 and 3, respectively. The different number of topics in each time period appears to confirm our decision to split the data set for topic modelling and is suggestive that the topics discussed broadened over time. We note, however, that while periods 2 and 3 have a similar number of documents (comments aggregated by parent post), period 1 is considerably smaller (see Section~\ref{sec:methods:topicmodel}). 



Table~\ref{tab:topics} shows which topics were present, their proportion and the topic ranking for each time period. 
In periods 1 and 3, the top-ranking topic was {\em Quitting}, 
whereas in period 2, when r/antiwork itself was being featured in numerous news stories, the top-ranking topic was {\em Reddit}.
The top-3 topics for all time periods were the same: {\em Quitting}, {\em Reddit} and {\em Mental Health} 
and accounted for 22.5-27.5\% of the content on r/antiwork.
In total, 17 topics appeared in all three time periods, accounting for 60.6-74.1\% of content.
Each time period had unique topics, many of which were based on seasonal events and major stories in the news media.
Period 1 included {\em Leisure} (i.e.~hobbies and free time) and {\em Social Security} (disability, welfare). 
Period 2 included 
{\em Holidays} (period 2 covered both Thanksgiving and Christmas), 
{\em Corporations} (related to, for example, Kellogg's union busting activities) and 
{\em Pandemic} (in particular, stories of working during the pandemic). 
Lastly, period 3 included topics for 
the {\em Fox News Interview}, 
{\em Working from Home} (in opposition to companies' post-pandemic return to office policies) and 
{\em Reproductive Rights} (related to the leaked U.S.~Supreme Court draft decision to overturn Roe v.~Wade). 
Topics confined to a single time period, however, tended to be relatively minor and were generally present in the long-tail of the topic distribution.

\section{Discussion}

Our study aimed to explore how user activity on r/antiwork was impacted by a gradual, sustained increase of subscribers, followed by a period of accelerated growth coinciding with mainstream media coverage of the Great Resignation. Instead, we found an online community where the parallel surge in posts and comments became decoupled from subscriber growth and collapsed after Doreen Ford's Fox News interview even as the number of subscribers continued to rise (RQ1). 
Change point detection provided suggestive evidence that user activity was driven by mainstream media events in mid-October 2021 and late January 2022, as the dates of these events were independently identified in both posting and commenting data (Figures~\ref{fig:posts} and \ref{fig:comments}, respectively).
We found that different types of users had a disproportionate influence on overall activity, with light posters and heavy commenters being responsible for almost a third of posts and comments, respectively (Figures~\ref{fig:posts} and \ref{fig:comments}) (RQ2). 
Light and heavy posters were responsible for similar proportions of posts prior to October 2021, but then gradually diverged until light posters were responsible for almost half of all posts by the end of July 2022 where our data set ends (Figure~\ref{fig:postsprop}). 
Commenting trends, on the other hand, appeared to be undisturbed by October's surge in new subscribers and January's collapse in activity (Figure~\ref{fig:commentsperpost}). While there was a spike in heavy commenters making their last comment immediately following the Fox News interview, it does not appear to have been a sufficient reduction to affect broader trends.
Lastly, despite anecdotal observations that the quality of discussion on r/antiwork had declined due to subscriber growth, we found no evidence to support this claim (RQ3). In general, the main topics of discussion were the same in all three time periods studied: the top-ranked topics were always {\em Quitting}, {\em Reddit} and {\em Mental Health}, and the topics shared by all time periods accounted for 60.6-74.1\% of content (Table~\ref{tab:topics}). Furthermore, we believe that we underestimated the degree of similarity between topic distributions, because a topic in one time period would sometimes correspond to two topics in another time period (e.g.~{\em Food/Drugs} in period 1 versus separate {\em Food} and {\em Drugs} topics found in both period 2 and 3).
%

%

Many studies use topic modelling to identify what is being discussed in online communities and to identify changes over time. 
We identified {\em Mental Health} and {\em Quitting} as two of the most prevalent topics on r/antiwork. 
This finding is in agreement with a study by del Rio-Chanona et al.~that identified mental health issues as one of the main reasons for members of the r/jobs subreddit to quit their jobs, especially since the onset of the pandemic \cite{del2022mental}. We also saw the introduction of a {\em Pandemic} topic that was unique to the period from October 25 2021-January 25 2022 (period 2), suggesting that the surge of new members shared their work-related experiences from during the COVID-19 pandemic. 
We found no evidence of a change in the topic distribution between time periods, in accordance with other studies of massive growth in online communities \cite{lin2017better}. This does not, however, discount the fact that an influx of newcomers could subtly change the feel of an online community. For example, Haq et al.~identify significant differences in writing style between new and long-term users, with new users writing shorter comments with more emojis \cite{Haq2022short}. Numerous studies also point to the temporary nature of long-term member grievances: Lin et al.~observed a dip in upvotes in newly defaulted subreddits that recovered quickly afterwards and, moreover, that complaints about low quality posts did not increase in frequency after defaulting \cite{lin2017better}. In an interview-based study, Kiene et al.~showed how r/nosleep attributed the subreddit's resilience in the face of sustained growth to active moderators and a shared sense of community \cite{kiene2016surviving}. It seems likely that this was the case with r/antiwork as well: several of the moderators have been involved since the subreddit's inception and have publicly championed the political objectives of the antiwork movement. Furthermore, there is a consistent hard core of heavy commenters, implying a sense of community at least among long-term members.

%

\subsection{Limitations}

In our study, we observed how mainstream media events coincided with changes in activity on r/antiwork (subscriber count, posting and commenting), but it is unclear to what extent these events were causal. In October 2021, the topics discussed on r/antiwork happened to align with the broader zeitgeist of worker dissatisfaction following the COVID-19 pandemic, so it seems likely that members would have found the subreddit through other means, such as the Reddit front page. Our findings related to the Fox News interview, however, do not appear to suffer from this limitation as the fallout had such wide and direct consequences, including the drop in posting and commenting activity, the loss of members, and was even captured by the topic model as a distinct topic.


Another limitation was the lack of data available from before 2019, when r/antiwork was a smaller and more focused community. Had we been able to include the earliest data, we might have seen greater differences in the topic distribution between then and 2022 than what we observed in our study. However, being a small sample, it would have had significant variance, leading to issues with the interpretation of results.

\subsection{Future Research}

In this study, we focused on characterising the development of r/antiwork and looked at how the surge of new members impacted user behaviour and what topics were being discussed. In future research, we plan to further investigate the aftermath of the Fox News interview. On January 26 2022, a subreddit called r/WorkReform was founded by disgruntled members of r/antiwork, gaining over 400,000 subscribers within 24 hours. We want to investigate the differences between the two subreddits in terms of users, topics discussed and reactions to major events, such as the overturning of Roe v.~Wade in June 2022.
Second, we want to take a deeper look at the users of r/antiwork, using their other activity on Reddit to understand why they behave the way they do. We believe that users who want to post about quitting their job could be more dissimilar to one another than heavy commenters whose interests are more likely to be focused on topics in r/antiwork.

\section{Conclusion}

In this paper, we presented an study of 
how subscribers, posts and comments on r/antiwork were impacted by events in the media,
how heavy and light user behaviour differed from one another, and 
a content analysis based on topic modelling to show how the discourse on the subreddit evolved.
We have shown that, despite the continuing rise of subscribers, activity on r/antiwork collapsed after the Fox News interview on January 25 2022. 
We showed that heavy commenters and light posters have a disproportionate influence on subreddit activity, making almost a third of overall comments and posts, respectively. Over time, light posters have become responsible for an increasing proportion of posts, reaching almost 50\% of posts by the end of July 2022. Heavy and light commenters, however, appeared unaffected by the surge of users, being responsible for approximately the same number of comments per post throughout the period studied. Commenting trends were not even impacted when 4.4\% of heavy commenters made their last comment on r/antiwork between January 26-28 2022 after the broadcast of the Fox News interview. Lastly, the influx of new users did not appear to change the topical content of discussion: all three time periods had the same top-3 topics: {\em Quitting}, {\em Reddit} and {\em Mental Health}. Each time period had distinct topics, but they tended to be related to seasonal events and ongoing developments in the news. Overall, we found no evidence of major shifts in the topical content of discussion over the period studied.
%


\balance
\bibliographystyle{ACM-Reference-Format}
\bibliography{sample-authordraft}


\appendix
\clearpage
\onecolumn

\section{Appendix}
\subsection{Topic models}

\renewcommand{\thetable}{A\arabic{table}}

\begin{table*}[h]
\centering
\resizebox{0.84\textwidth}{!}{ 
\begin{tabular}{l|l|c|c|c|c|c|c}
                    &                                  & \multicolumn{2}{|c|}{Period 1} & \multicolumn{2}{c|}{Period 2} & \multicolumn{2}{c}{Period 3} \\
Topic name          & Example keywords                 & Proportion & Rank & Proportion & Rank & Proportion & Rank \\ \midrule
{\bf Quitting}            & boss, quit, notice               & 0.102      & 1    & 0.075      & 3    & 0.084       & 1    \\
{\bf Reddit}              & sub, post, comment               & 0.087      & 2    & 0.087      & 1    & 0.062      & 3    \\
{\bf Mental Health}       & better, help, depression         & 0.086      & 3    & 0.085      & 2    & 0.079       & 2    \\
Leisure             & life, enjoy, hobby               & 0.081      & 4    &            &      &            &      \\
{\bf Economics}           & capitalism, ubi, socialism       & 0.069      & 5    & 0.068      & 4    & 0.036      & 9    \\
Reactions                 & shit, fuck, lol                  & 0.062      & 6    &            &      &            &      \\
{\bf Scheduling}          & shift, break, overtime           & 0.061      & 7    & 0.026      & 14   & 0.038      & 8    \\
{\bf Wages}               & minimum, wage, paid              & 0.047      & 8    & 0.043      & 10   & 0.030      & 11   \\
{\bf Hiring}              & interview, resume, hire          & 0.040      & 9    & 0.050      & 7    & 0.055      & 5    \\
{\bf Money}               & money, tax, wealth               & 0.037      & 10   & 0.044      & 9    & 0.032      & 10   \\
Society                   & society, life, shelter        & 0.033      & 11   &            &      & 0.042      & 7    \\
{\bf Family}              & family, kid, parent              & 0.032      & 12   & 0.017      & 21   & 0.020      & 19   \\
{\bf Labour Rights}       & union, law, employee             & 0.031      & 13   & 0.048      & 8    & 0.052      & 6    \\
Sick Leave          & sick, covid, leave               & 0.031      & 14   &            &      & 0.022      & 16   \\
{\bf Finance}             & money, stock, debt               & 0.029      & 15   & 0.023      & 16   & 0.019      & 21   \\
{\bf Retail Sector}       & store, customer, retail          & 0.028      & 16   & 0.054      & 6    & 0.015      & 28   \\
Qualifications      & degree, college, school          & 0.028      & 17   & 0.026      & 13   &            &      \\
{\bf Location}            & country, america, europe         & 0.027      & 18   & 0.018      & 18   & 0.006      & 35   \\
{\bf Service Sector}      & food, restaurant, tip            & 0.021      & 19   & 0.016      & 24   & 0.023      & 15   \\
{\bf Housing}             & house, rent, landlord            & 0.019      & 20   & 0.026      & 12   & 0.018      & 24   \\
{\bf Politics}            & right, vote, party               & 0.015      & 21   & 0.025      & 15   & 0.016      & 26   \\
Food/Drugs          & food, drug, weed                 & 0.011      & 22   &            &      &            &      \\
{\bf Education}           & school, teacher, class           & 0.010      & 23   & 0.013      & 28   & 0.021      & 17   \\
Transportation      & car, drive, gas                  & 0.009      & 24   &            &      & 0.015      & 27   \\
Social Security     & disability, benefit, welfare     & 0.002      & 25   &            &      &            &      \\
Generic Work        & company, employee, manager       &            &      & 0.063      & 5    & 0.008      & 33   \\
Work Benefits       & hr, vacation, pto                &            &      & 0.038      & 11   &            &      \\
Corporations        & amazon, walmart, kelloggs        &            &      & 0.022      & 17   &            &      \\
Healthcare          & insurance, healthcare, doctor    &            &      & 0.018      & 19   & 0.019      & 22   \\
Pandemic            & covid, mask, vaccine             &            &      & 0.018      & 20   &            &      \\
Boomers             & boomers, old, retirement         &            &      & 0.017      & 22   & 0.014      & 29   \\
Food                & food, eat, coffee                &            &      & 0.016      & 23   & 0.013      & 31   \\
Holidays            & holiday, christmas, thanksgiving &            &      & 0.016      & 25   &            &      \\
Unions              & union, strike, worker            &            &      & 0.016      & 26   & 0.017      & 25   \\
Police              & police, cop, crime               &            &      & 0.014      & 27   & 0.013      & 30   \\
Discrimination      & racist, discrimination, gender   &            &      & 0.012      & 29   & 0.013      & 32   \\
Drugs               & drug, test, weed                 &            &      & 0.004      & 30   & 0.005      & 39   \\
Improved Wages      & raise, offer, bonus              &            &      &            &      & 0.058      & 4    \\
Fox News Interview  & antiwork, fox, interview         &            &      &            &      & 0.026      & 12   \\
Working from Home   & home, wfh, remote                &            &      &            &      & 0.024      & 13   \\
Bathroom Breaks     & bathroom, break, toilet          &            &      &            &      & 0.024      & 14   \\
Health Sector Work  & doctor, nurse, patient           &            &      &            &      & 0.020      & 18   \\
Communications      & phone, email, text               &            &      &            &      & 0.019      & 20   \\
Elon Musk           & elon, tesla, billionaire         &            &      &            &      & 0.018      & 23   \\
Reproductive Rights & woman, baby, abortion            &            &      &            &      & 0.007      & 34   \\
Dress Code          & hair, uniform, tattoo            &            &      &            &      & 0.006      & 36   \\
Driving Work        & pizza, delivery, uber            &            &      &            &      & 0.005      & 37   \\
Community           & community, local, volunteer      &            &      &            &      & 0.005      & 38   \\
Scams               & scam, crypto, hacker             &            &      &            &      & 0.002      & 40  \\ \bottomrule
\end{tabular}
}
\caption{Labelled topics from three time periods ordered by prevalence in the first time period they appear in.
Bold topic names are present in all three time periods. Example keywords are not necessarily the highest probability 
words, but representative of the topic across all time periods it appears in. Proportion columns may not add up to 1.0 due to rounding.}
\label{tab:topics}
\end{table*}

\end{document}